# Polarization-Independent and High-Efficiency Dielectric Metasurfaces Spanning 600-800 nm Wavelengths

Qi-Tong Li[1,*], Fengliang Dong[2,*], Bo Wang[1], Fengyuan Gan[1], Jianjun Chen[1,3], Weiguo Chu[2], Yun-Feng Xiao[1,3], Qihuang Gong[1,3], and Yan Li[1,3]

[1]State Key Laboratory for Mesoscopic Physics, Department of Physics, Peking University and Collaborative Innovation Center of Quantum Matter, Beijing 100871, China.

[2]Nanofabrication Laboratory, National Center for Nanoscience and Technology, Beijing 100190, China.

[3]Collaborative Innovation Center of Extreme Optics, Shanxi University, Taiyuan, Shanxi 030006, China

* These authors contributed equally to this work. Correspondence and requests for materials should be addressed to W.C. (email: wgchu@nanoctr.cn) and Y.L. (email: li@pku.edu.cn).

**Abstract**

Artificial metasurfaces are capable of completely manipulating the phase, amplitude, and polarization of light with high spatial resolutions. The emerging design based on high-index and low-loss dielectrics has led to the realization of novel metasurfaces with high transmissions, but these devices usually operate at the limited bandwidth, and are sensitive to the incident polarization. Here, for the first time we report experimentally the polarization-independent and high-efficiency dielectric metasurfaces spanning the visible wavelengths about 200 nm, which are of importance for novel flat optical devices operating over a broad spectrum. The diffraction efficiencies of the gradient metasurfaces consisting of the multi-fold symmetric nano-crystalline silicon nanopillars are up to 93% at 670 nm, and exceed 75% at the wavelengths from 600 to 800 nm for the two orthogonally polarized incidences. These dielectric metasurfaces hold great potential to replace prisms, lenses and other conventional optical elements.

Since the era of Huygens, reshaping the wavefront of the light has been depending on gradual phase changes from the light propagation in different materials with elaborately tailored geometries, and various conventional optical devices have been developed. However, their complex spatially curved surfaces and considerable thicknesses seriously imped their applications to integrated optical systems and the arbitrary wavefront control. Fortunately, metasurfaces, or two-dimensional metamaterials, can afford to circumvent the disadvantages by introducing nano-resonator (meta-atom) arrays in the interface of two media[1]. A specially designed metallic plasmonic resonator can bring forth a specific phase delay for the scattering light, which is determined by its resonance modes that can be tailored by changing the physical sizes of the meta-atoms. As a result, an abrupt phase discontinuity is obtained in a plasmonic metasurface to allow directly manipulating the wave fronts of the scattering light. Many novel applications based on metallic metasurfaces have been demonstrated, including the anomalous refraction and reflection[1-3], beam shaping[4-6], transformation between propagating waves and surface waves[7], interaction of photonic spin and orbital momentum[8] and holograms[9, 10].

Nevertheless, practical applications of plasmonic metasurfaces are, to great extent, restricted by the extremely low efficiencies caused by their intrinsically high ohmic loss and the finite scattering cross sections. To address this issue, a trilayer structure is proposed to ensure the perfect performance in reflection mode[11-19] and extended to the successful phase control in transmission mode with a relatively high efficiency in the terahertz range[16].

A promising path to solve this problem is to use dielectric metasurfaces consisting of monolayer dielectric resonators made of high refractive index materials[20, 21]. Based on the Mie scattering theory, low-loss dielectric metasurfaces can provide much stronger forth scattering by exciting both electric and magnetic resonances simultaneously[22,-24], which enables the effective phase manipulation for the full electric field with high transmittance[25-25, 29, 33]. Furthermore, such a single transmission mode avoids the post-selection of the polarization and favors the design of polarization-independent metasurfaces [28, 29].

Here we both theoretically and experimentally demonstrate the realization of polarization-independent broadband dielectric metasurfaces at visible wavelengths, which would boost the practical applications of meta-devices. We show that all the meta-atoms with a multi-fold symmetry have a polarization-independent response. Such a high flexibility renders it possible to design metasurfaces with additional amazing properties. The broadband feature (550-800nm) is achieved by introducing a $3\pi$ phase control. Although the phase delay of an individual meta-atom inevitably changes with the wavelength of the incident light, the difference of the phase delays from two specially designed meta-atoms can remain constant. To exhibit these advantages, a gradient dielectric metasurface is designed and fabricated as shown in Figure 1. Each super cell consists of six meta-atoms (nano-resonators). The diffraction efficiency is as high as 93% in visible frequencies, demonstrating a great potential to form many novel flat optical devices such as prisms, lenses, beam generators and holograms.

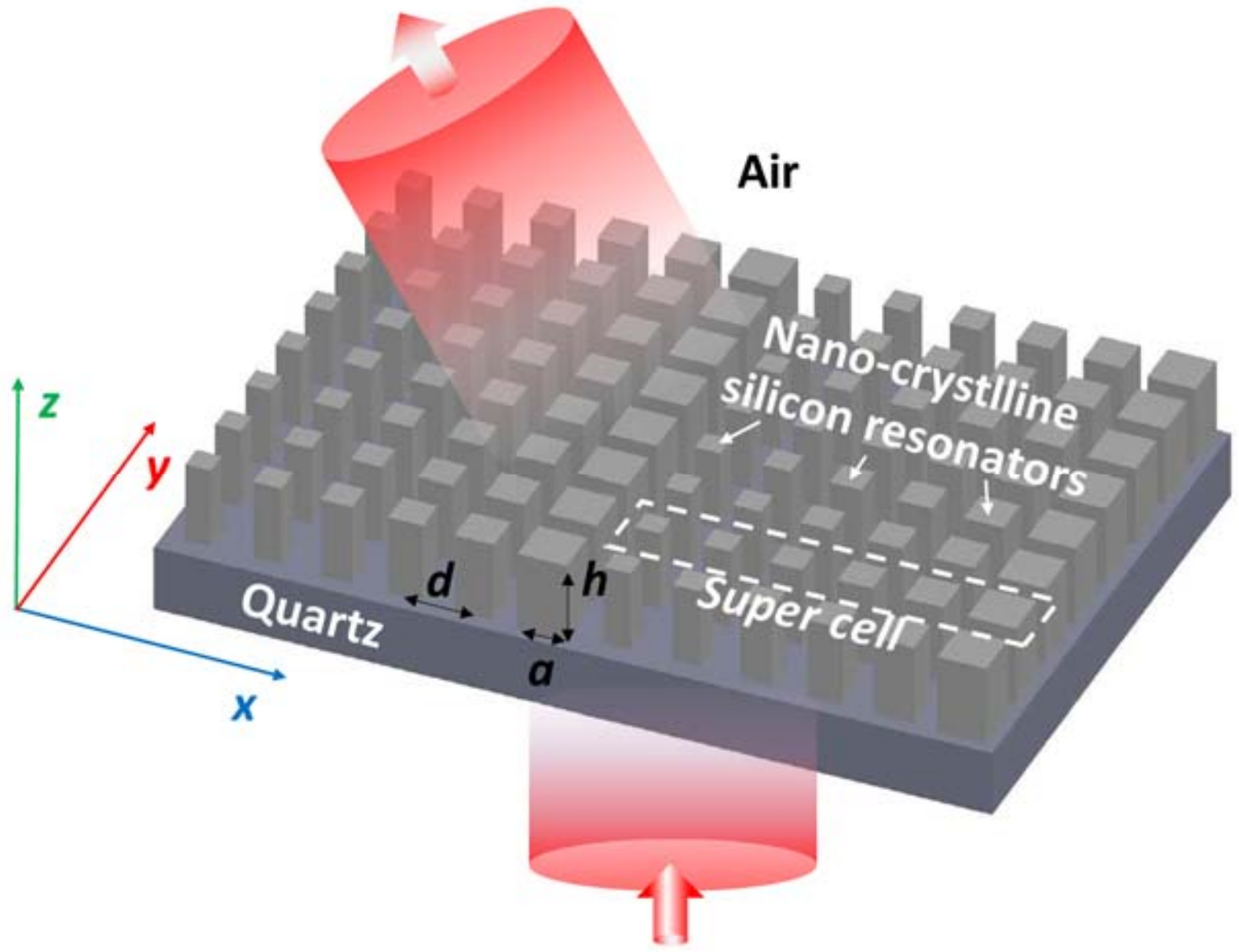

**Figure 1 | Schematic of a gradient metasurface.** An array of nano-crystalline silicon nano-resonators is arranged on a quartz substrate and the beam is normally incident along $z$-axis. $h$, $a$ and $d$ are the height, length and lattice constant of the dielectric resonators, respectively. A super cell includes six different resonators.

## Results

**Design of Polarization-independence.** We first consider a simple infinite metasurface consisting of the same meta-atoms to illustrate how to realize the polarization-independence. We focus on the influence of the geometry of meta-atoms, and temporarily neglect the Bravais lattice aberration. In this case, rotating the entire infinite metasurface by an angle of $\theta$ is presumed to be equivalent to individually rotating each meta-atom by the same angle. Assuming that a right circular polarization plane wave is normally incident along $z$ axis, the incident electric field is $\boldsymbol{E_i}(z,t) = \boldsymbol{E_i}(z)e^{-i\omega t} = \mathrm{E}_0(\hat{e}_x + i\hat{e}_y)e^{ikz-i\omega t} = E_0(z)|\boldsymbol{R}\rangle e^{-i\omega t}$. Since the dimension of the meta-atom is smaller than the operating wavelength, the infinite metasurface can be regarded as a homogeneous 2D material, with the transmission electric field $\boldsymbol{E_t}$ described by

$$\boldsymbol{E_t}(x,y,z,t) = \boldsymbol{E_t}(z)e^{-i\omega t} = [\mathrm{E_R}(z)|\boldsymbol{R}\rangle + \mathrm{E_L}(z)|\boldsymbol{L}\rangle]e^{-i\omega t} \tag{1}$$

where $\mathrm{E_R}(z)$ and $\mathrm{E_L}(z)$ in Eq.(1) are complex amplitudes of the right-circular polarization ($|\boldsymbol{R}\rangle$) and the left-circular polarization ($|\boldsymbol{L}\rangle$), respectively. $\hat{e}_x$ and $\hat{e}_y$ are unit vectors along $x$- and $y$-directions. Due to the Pancharatnam-Berry phase[30], rotating the metasurface by $\theta$ will induce a $2\theta$ phase change in the cross polarization component with respect to that of the incident light. Therefore, the transmission electric field satisfies

$$\boldsymbol{E^\theta}(z) = \mathrm{E_R}(z)\big|\boldsymbol{R}\rangle + \mathrm{E_L}(z)e^{i2\theta}\big|\boldsymbol{L}\rangle \tag{2}$$

We should note that such a phenomenon is fundamentally attributed to the isotropic properties of the circularly polarized plane wave in the $x-y$ plane. The analytical solution of Eq. (2) does not change with the rotation operation in the local coordinate fixed in the infinite metasurface. This additional Pancharatnam-Berry phase is a purely spatial geometry effect, which holds for all the meta-atoms. When the meta-atom has an $n$-fold symmetry, rotating each meta-atom by $2\pi/n$ does not change the metasurface, and its transmission electric field is the same as that before operation

$$\boldsymbol{E^{2\pi/n}}(z) = \boldsymbol{E}(z) = \mathrm{E_R}(z)|\boldsymbol{R}\rangle + \mathrm{E_L}(z)\,|\boldsymbol{L}\rangle \tag{3}$$

On the other hand, this rotation leads to the Pancharatnam-Berry phase, therefore

$$\boldsymbol{E^{2\pi/n}}(z) = \mathrm{E_R}(z)|\boldsymbol{R}\rangle + \mathrm{E_L}(z)e^{\frac{i4\pi}{n}}|\boldsymbol{L}\rangle \tag{4}$$

From Eqs. (3) and (4), we get the relationship

$$\mathrm{E_L}(z) = \mathrm{E_L}(z)e^{i4\pi/n} \tag{5}$$

When $n$ is 1 or 2, the Eq.(5) is automatically satisfied. However, when the integer $n$ is larger than 2, the equation is valid only if $\mathrm{E_L}(z) = 0$. Similarly, for a left circularly polarized incident plane wave, $\mathrm{E_R}(z) = 0$ is also the only feasible solution. Therefore, for resonators with a multi-fold symmetry ($n$>2), a cross-polarized transmission light cannot be excited. The more rigorous analysis is presented in the Supplementary information. Finally, in the horizontal and vertical polarization frame using $\hat{e}_x$ and $\hat{e}_y$ as the unit vectors, the transmission matrix $\mathbf{A}$ should be rewritten as

$$\mathbf{A} = \mathbf{A}_{xy} = \begin{bmatrix} t & 0 \\ 0 & t \end{bmatrix} = t\,\mathbf{I} \tag{6}$$

As shown in Eq. (6), the *x* and *y* components of the incident plane wave not only transmit the metasurface independently but also have the same amplitude and phase delay after the metasurface. Similar results have been demonstrated recently in controlling harmonic generations in nonlinear materials[31-33].

Furthermore, if the Bravais lattice of the metasurface has a $m$-fold symmetry with *m* being a multiple of *n*, the isotropic response should remain since a rotation operation can reproduce the original metasurface without any aberration. However, a slight deviation would occur as long as *m* is not a multiple of *n*. In this case, the rotation of the entire infinite metasurface by $\theta$ is not exactly equivalent to the rotation of each meta-atom by the same angle. To verify our analysis, numerical simulations are performed using the commercial software COMSOL. In calculation, the wavelength of the incident light is fixed at 632.8nm from a He-Ne laser. The refractive index of the resonators and the quartz substrate are set to be 3.63+0.06*i* and 1.5, respectively. The height of the resonators is 370nm according to our experimental results. Figs. 2a-2d show both the phase delay and the amplitude of the transmission electric field in two orthogonal polarizations as a function of the resonator dimension. The simulated data are in good agreement with our prediction. In Fig. 2a (square resonators (*n*=4) in a square lattice (*m*=4)) and Fig. 2c (regular triangle resonators (*n*=3) in a hexagonal lattice (*m*=6)), the two curves for *x*- and *y*-polarizations perfectly overlap. At the same time, the expected slight differences are observed in Fig. 2b (square resonators (*n*=4) in a rectangular lattice (*m*=2)) and Fig. 2d (regular triangle resonators (*n*=3) in a square lattice (*m*=4)) because of the mismatch of the symmetries of resonators and lattices.

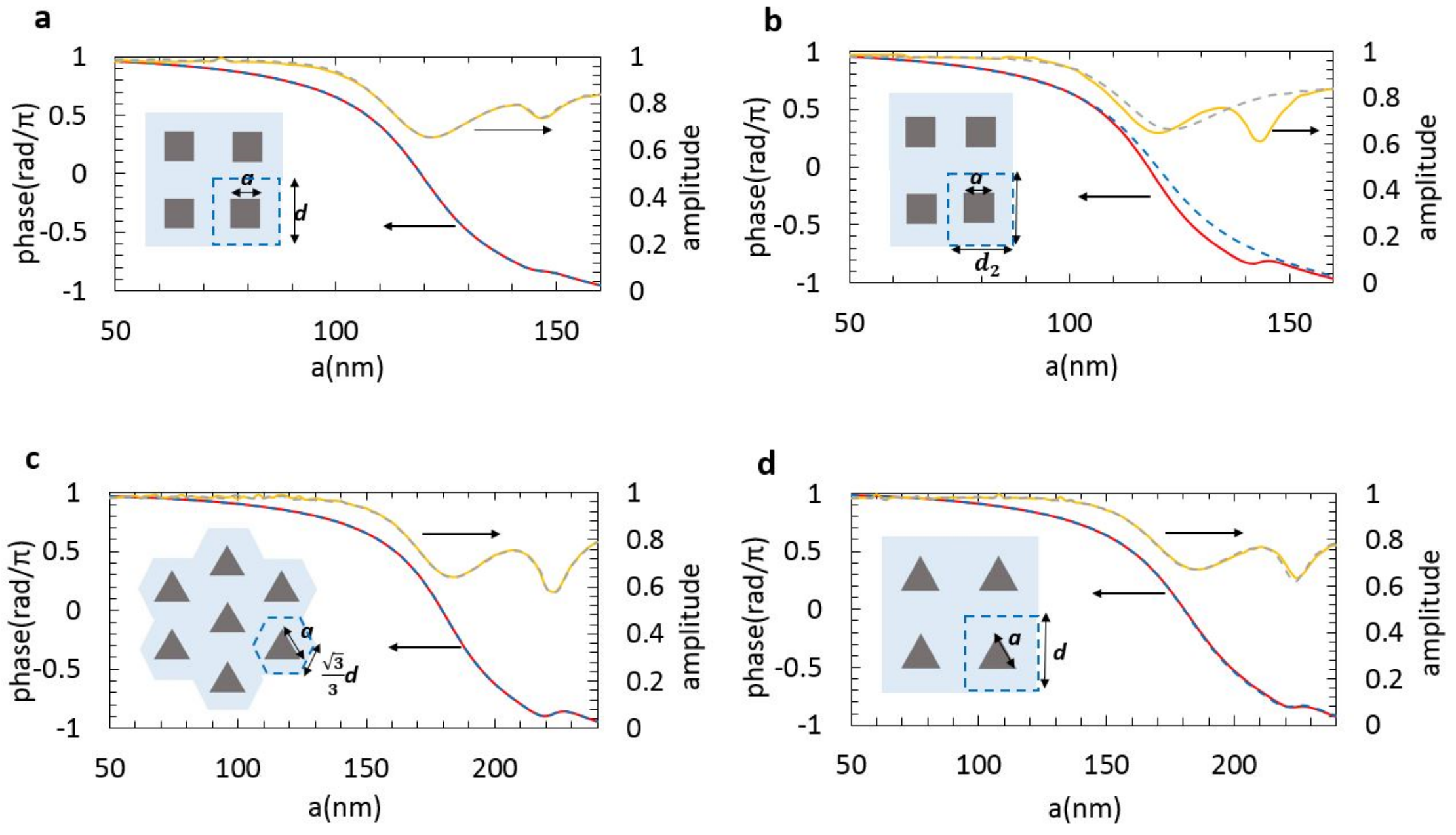

**Figure 2 | Numerical simulations of the phase delay and the amplitude of the transmission light through an infinite metasurface as a function of the size of meta-atoms.** The red solid line, the blue dash line, the yellow solid line and the gray dash line represent the phase delay of x-polarization, the phase delay of y-polarization, the amplitude of x-polarization, respectively. The amplitude is normalized to the intensity of incident light. **(a), (b)** Meta-atoms have a square cross section with a square lattice and a rectangular ($d_1 = 1.1d, d_2 = 0.9d$) lattice, respectively. The length *a* is changing from 50nm to 160nm, and *d*=300nm is the lattice constant. **(c), (d)** Meta-atoms have a regular triangle cross section with a hexagonal lattice and a square lattice, respectively. The length *a* is changing from 50nm to 240nm, and *d*=200$\sqrt{3}$nm is the lattice constant.

**Broadband at the visible wavelengths.** The polarization-independent metasurfaces can be surprisingly realized only by tailoring the symmetry of constituent resonators and the Bravais lattice instead of the dimensions of resonators as demonstrated above. More importantly, this also leads freedom to achieve the amazing broadband at the visible wavelengths with a full $2\pi$ phase control, which is exceptionally beneficial for spatial phase modulation optical elements such as prisms and lenses. As an example, we analyze the metasurface consisting of square resonators in a square lattice as shown in Fig. 2a, which satisfies the polarization-independent conditions aforementioned. The resonators are made of the nano-crystalline silicon due to its pertinent refractive index and relatively low loss in visible frequencies as presented in Supplementary Fig. 1. The lattice constant $d$ is set to be 300nm (about the half of the wavelength of the incident light) to avoid unwanted diffractions. It has been recently demonstrated that the perfect match of the electric and magnetic resonances results in a reflectionless effect thanks to the constructive interference for forward scattering light and the destructive interference for backscattering light[34-36]. Here we set the height $h$ of the resonators to 370nm for two reasons: it not only ensures the relatively high transmittance of dielectric resonators in a broadband range by exciting electric type and magnetic type resonances simultaneously; more importantly, it achieves a smooth phase gradient to ease the harsh fabrication conditions encountered by the visible-frequency meta-devices.

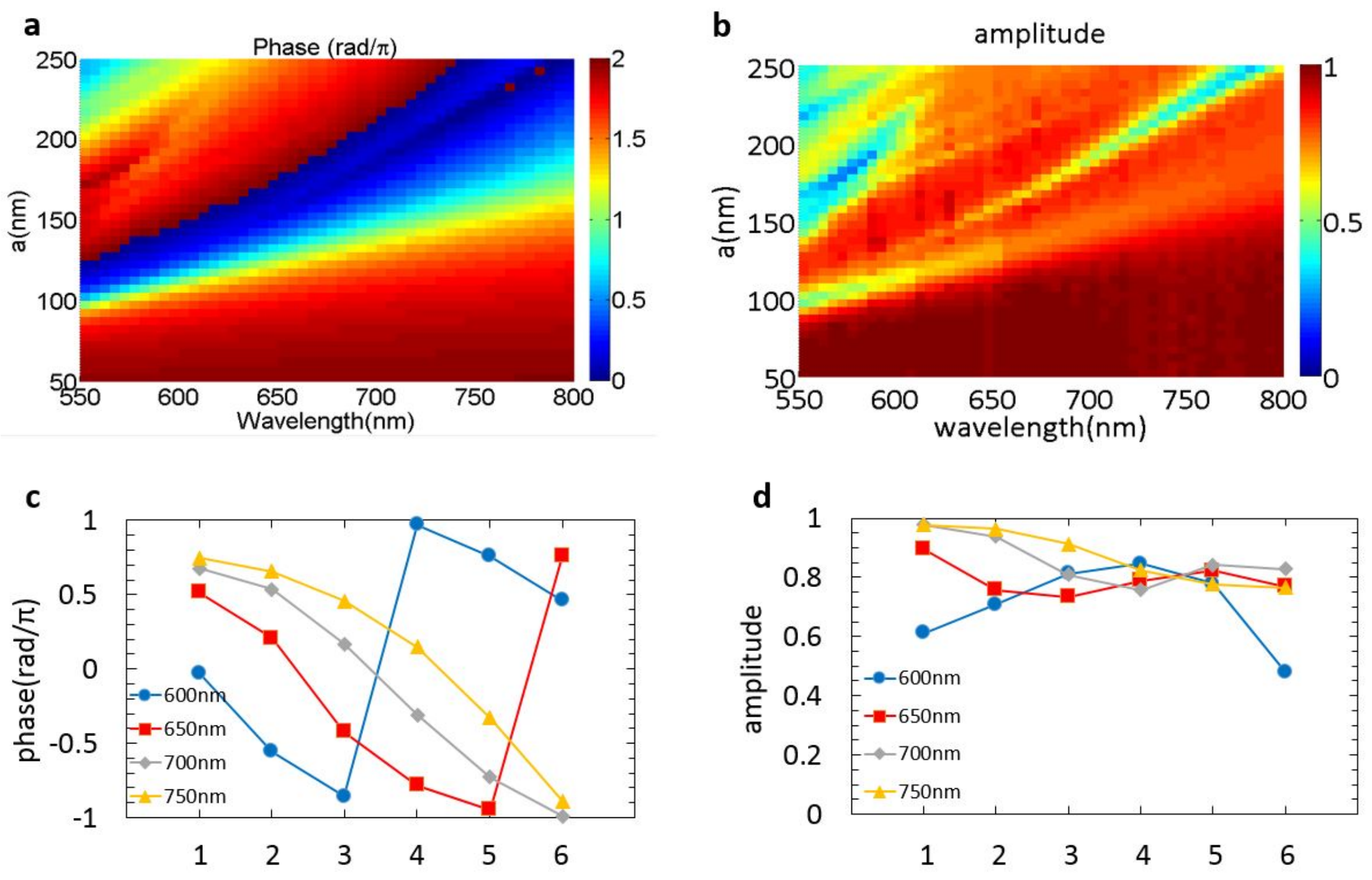

**Figure 3 | Simulated results. (a), (b)** Numerical simulations of phase delay and amplitude of transmission light with an infinite metasurface as a function of wavelength from 550nm to 800nm and the length of meta-atoms from 50nm to 250nm. **(c), (d)** the phase delay and the amplitude of transmission light through the six chosen meta-atoms at wavelength of 600nm (blue circle), 650nm (orange square), 700nm (gray diamond) and 750nm (yellow triangle).

We find in Fig. 3a that by changing the dimensions of resonators, not only a $2\pi$ but also a $3\pi$ phase control can be realized at short wavelengths. It may be trivial for a monochromatic device because only the difference of the phases other than phases themselves plays a key role. However, the $3\pi$ phase control indeed gives rise to the broadband feature. Although the phases change with the wavelength of the incident light for a dimension fixed resonator array due to different resonance excitations in different frequencies, we can keep the phase difference constant for resonators with different sizes (i.e. the chosen meta-atoms as the basic units to compose the metasurface) in a broad spectrum by setting the phase delay from $0\pi$ to $2\pi$ at the longer wavelength and from $\pi$ to $3\pi$ at the shorter wavelength. The simulations in Fig. 3a verify its feasibility. When the size $a$ ranges from 100nm to 250nm, the phase delay changes from $\pi$ to $3\pi$ at 550nm but from $0\pi$ to $2\pi$ at 800nm. At the same time, the amplitude of the transmission light shown in Fig. 3b is still relatively high and fluctuates only slightly from 600 to 800nm. Figs. 3c and 3d present the phase delays and amplitudes of the transmission lights for six constituent resonators to compose a super cell as shown in Fig.1. The dimensions of the resonators labeled by 1 to 6 are 110, 120, 135, 150, 170 and 210nm, respectively. The lattice constant $d$ is 300nm. It is perspicuous that the phase delays of the six chosen meta-atoms uninterruptedly cover a range of $2\pi$ from 600 to 750nm with high transmissions, which makes the fabricated metasurface achieve a high diffraction efficiency in this wavelength range.

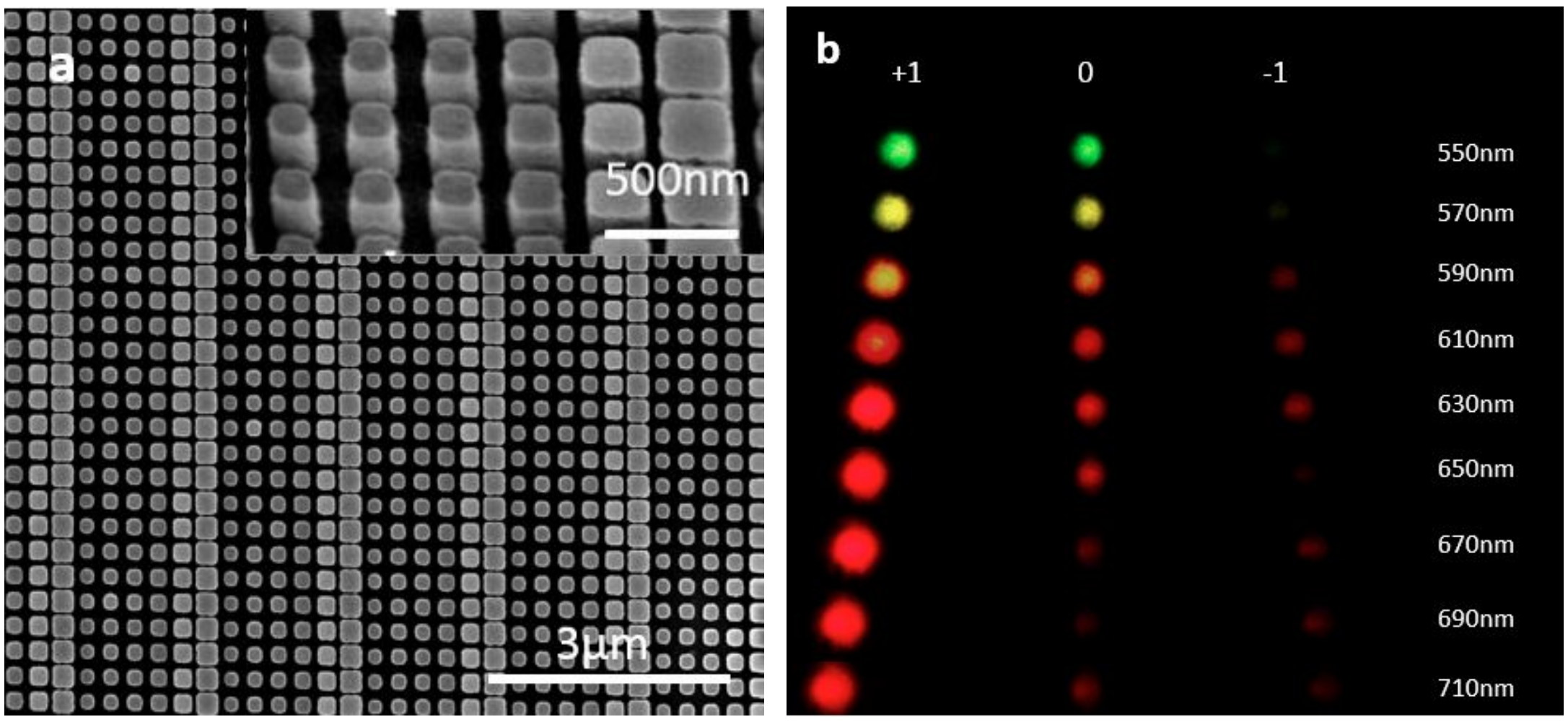

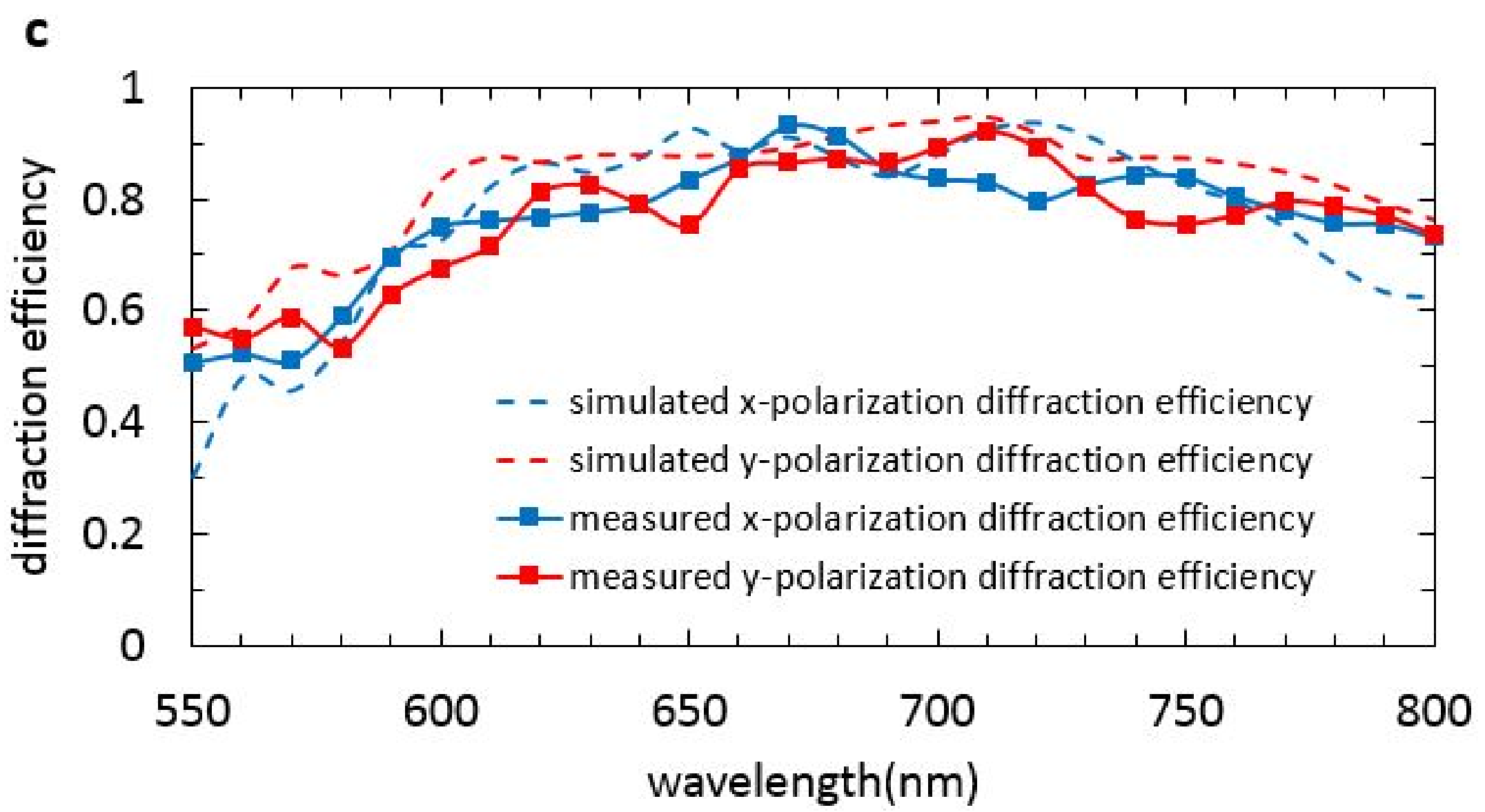

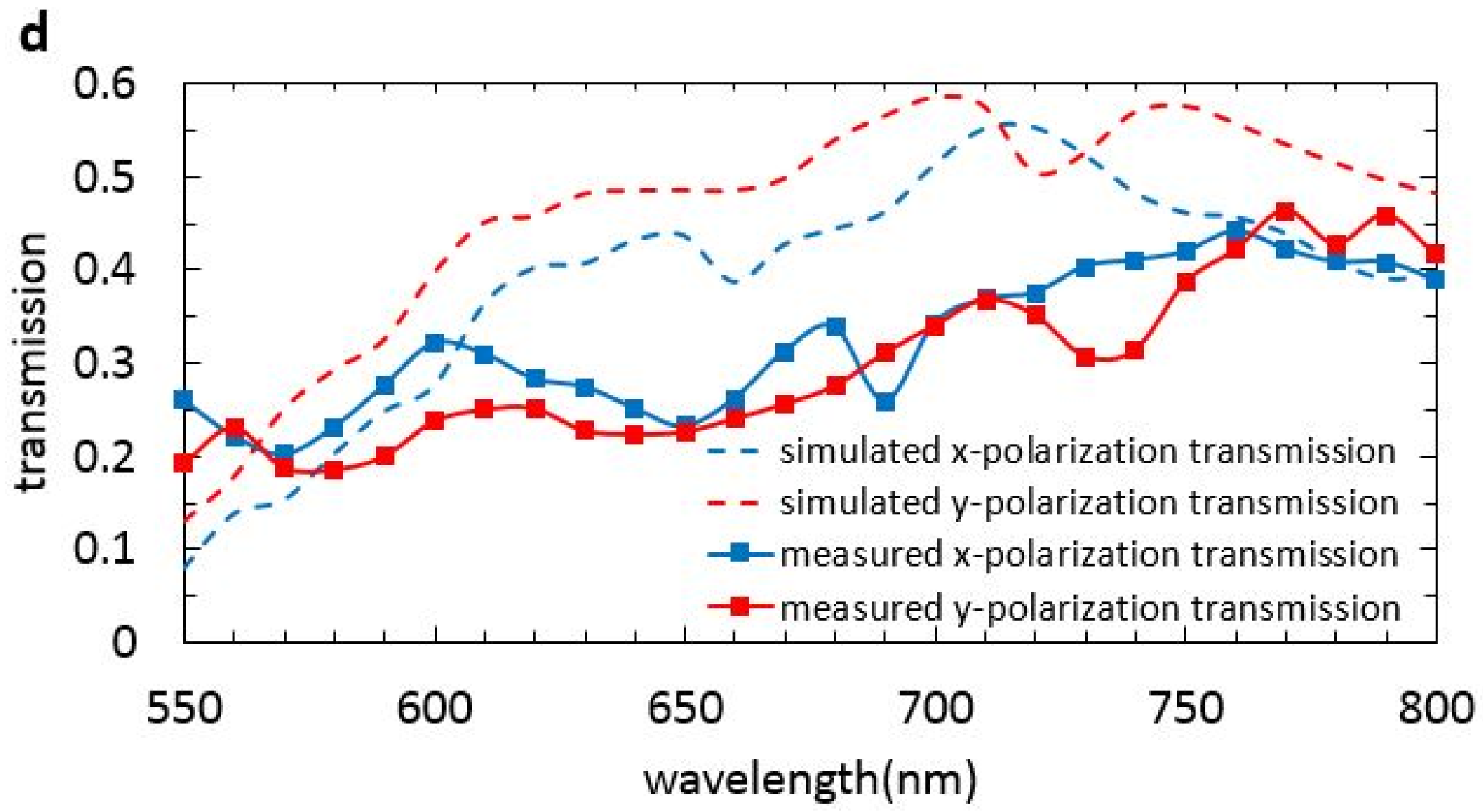

**Figure 4 | Experimental results. (a)** Top-view and titled-view SEM images of the sample. The dimensions of the six chosen meta-atoms are 110nm, 120nm, 135nm, 150nm, 170nm and 210nm, respectively. **(b)** Photos of the transmission light distribution at the wavelength of 550nm to 710nm. **(c)** Measured and simulated diffraction efficiency (defined as the ratio of the power of anomalous refraction beam (+1 order) and the sum power of the +1, 0, and -1 orders transmission light) spectra from 550nm to 800nm with x- and y-polarized incident light, respectively. **(d)** Measured and simulated transmittance (defined as the ratio of the power of anomalous refraction beam and the power of incident beam) spectra from 550nm to 800nm with *x*- and *y*-polarized incident light, respectively.

**Fabrication and characterization of the dielectric metasurfaces.** According to our designs as shown in Fig. 1, we fabricated a gradient dielectric metasurface from a 370nm-thick nano-crystalline silicon film on a quartz substrate using the procedure described in the Methods section. Each super-cell is composed of six resonators arrayed from small to large in dimension. The scanning electronic microscope (SEM) images in Fig. 4a revealed that the configuration of the fabricated metasurface agreed very well with our design. The periodicity also induces weak high-order diffractions. The refraction angle can be calculated by the formula $\theta_r = \arcsin(\lambda/6d)$, which is from 17.8° to 26.4° in the wavelength range from 550nm to 800nm. We measured all the intensities of the anomalous refraction beam (+1st order diffraction), the 0th order normal transmission beam, the -1st order diffraction beam and the incident beam from 550 to 800nm. Both the $x$- and y-polarized incident beams were adopted to verify the polarization-independence. To visualize the results, a series of photos in Fig. 4b were taken to show the intensity distributions of transmission lights. From the images, the refraction angles do increase with the wavelength as expected. Figure 4c shows the measured diffraction efficiencies from 550 to 800nm, which are amazingly higher than 75% from 600 to 800nm. The peak efficiencies, 93% and 92%, are at 670nm and 710 nm for the $x$- and $y$- polarized incident beams, respectively. The experimental results agree well with the simulations. Such salient broadband is much larger than the reported values[20, 25-27, 29, 38]. In Ref. 20, the wavelength range is about 50nm with the efficiencies higher than 50% and a peak diffraction efficiency of about 75%. In addition, the transmittances are as high as 20% to 45% in the wavelength range of 550-800nm as shown in Fig. 4d, which conform to the simulations. They are much higher than the theoretical limit of 25% for ultrathin metasurfaces (the height of the meta-atoms is much smaller than the wavelength of the incident light)[37]. The transmittances of 44% and 46% are observed at 760nm and 770nm for the $x$ – and $y$ - polarized incident beam, which are quite comparable to the recently reported light bending meta-devices such as 45% at 710nm in Ref. 27, 36% at 1550nm in Ref. 38, and 20% at 1800nm using a metallic trilayer structure in Ref. 39. It should be noted that the 46% transmittance is absolutely not the upper limit. By exploring new materials with lower loss in the visible frequencies, much higher transmittance can be expected. A recently published article[26] reported a 77% transmission for the anomalous refraction at 915nm where their materials exhibit a zero imaginary part of refractive index, which significantly paves a promising way to promote the transmission further. The combination of the polarization independence, the broadband, the high diffraction efficiency and the high transmittance in the visible frequencies can greatly facilitate the design, fabrication and application of higher performance optical elements such as prisms and lenses.

## Discussion

The slight aberrations between two orthogonal polarizations were observed in Figs. 4c and 4d. In fact, practical metasurfaces can hardly achieve a perfect polarization-independence even if the theoretical conditions are satisfied and the fabrication aberrations are reduced to infinitesimal. The intrinsic reason is that practical metasurfaces are surely composed of different types of meta-atoms to provide the arbitrary spatial phase modulation. Unlike the ideal periodic settings in simulations, the meta-atoms surrounding one specific meta-atom are different, and the resultant aberrations cannot be completely removed by improving the fabrication processing. As mentioned above, one of the feasible ways to further improve the polarization-independence is to replace the standard basic units by specifically tailoring spatially dependent meta-atoms according to their surrounding environments. Furthermore, if the light distribution also has a multi-fold symmetry, the polarization-independence can be guaranteed by the geometry symmetry of the light.

From Figs. 3b and 3d, our simulated transmission seems lower than those of some reflectionless devices[27, 29, 38], whereas the transmissions of the fabricated devices are actually comparable. The amazing results are attributed to the smooth variation of both the phase retardations and the amplitude of the transmitted light for our high quality metasurface, which significantly reduce the influence of both fabrication deviations and intrinsic aberrations by the destruction of the periodic conditions. In addition, the high quality of the metasurface here can also be ascribed to the big enough step sizes chosen for the six meta-atoms such as 10nm, 15nm, 15nm, 20nm and 40nm which can be fairly readily processed by both electron beam lithography and plasma etch. Even in the visible-frequency, the designed meta-device is also processible. It is worth noting that by optimizing the designs (say, using cross shape resonators) to eliminate the excursion of the difference of phases between two different meta-atoms, we can further improve linear responses of the meta-atoms to further broaden the operating band to cover the full visible wavelengths.

In conclusion, we both theoretically and experimentally demonstrate the first realization of polarization-independent broadband dielectric metasurfaces with high efficiencies for visible light. The polarization-independence originates from the proper symmetries of resonators and arrays, and the broad bandwidth is achieved by choosing suitable materials and physical sizes for resonators. Our metasurfaces highly improve the phase manipulation in transmission mode and possess the great potential to replace most of the conventional phase modulation optical elements such as lenses, prisms and phase plates by flat ones upon designing novel integrated photonic systems for imaging, communications and information processing.

## Methods

**Sample fabrication.** We fabricated the devices on a quartz substrate. The process began with the deposition of a 370nm-thick intrinsic nano-crystalline silicon film using an Inductively Coupled Plasma Enhanced Chemical Vapor Deposition System (ICPECVD, Sentech SI 500D). A 50-nm thick aluminum film was then deposited via electron beam evaporation, used as a charge-dissipation layer and hard mask. The 300nm thick positive electron beam resist (ZEP-520A) was coated and patterned using electron beam lithography (EBL, Vistec EBPG 5000+). The patterns were transferred into the aluminum and Si layer by subsequent etching using an Inductively Coupled Plasma (ICP, Sentech PTSA SI 500) etcher. Finally, the aluminum layer was removed at room temperature using aluminum etchant.

**Measurement procedure.** The measurement setup is schematically shown in Supplementary Fig. 2. Firstly, a supercontinuum laser (Fianium SC400-4) generated a linearly polarized Gaussian beam in the range of 550nm to 800nm. After the first polarizer to select the initial linear polarization, a quarter-wave plate was applied to change the linear polarization into the elliptical polarization. Then either *x*-polarized or *y*-polarized beam was selected by the second polarizer. The photos of the light distributions on a white screen were taken by a camera (Canon 5D Mark Ⅱ). The intensities of the transmission beam were measured by a power meter (Thorlabs PM100D).

**Simulations.** We performed the simulations in frequency domain by commercial software COMSOL. Using ports along *z*-axis and perfect matched layer, we characterize the phase delay and the amplitude of transmission electromagnetic fields. Both a normally incident plane wave and periodic conditions were assumed to save the computation time. However, we should concede that our simulation results, used as a local response, actually were derived under the conditions of an infinite metasurface with an array composed of the same resonators, which indeed may have some differences from our experimental conditions. To justify our simulation methods, we conducted full-field simulations for the whole super cells, and the results have a reasonable agreement with our expectations (Supplementary Fig. 3).

## Acknowledgements

This work was supported by the National Basic Research Program of China under Grant No. 2013CB921904, the National Natural Science Foundation of China under Grant Nos. 11474010 and 11134001, the Scientific Research Foundation for the Returned Overseas Chinese Scholars, State Education Ministry under Grant No.Y5691I11GJ, and Youth Innovation Promotion Association CAS under Grant No. Y5442912ZX.

## Author contributions

Q.L., F.D. and Y.L. conceived the idea; Q.L. performed the designs, simulations, and measurements; F.D. and W.C. fabricated the metasurfaces and assisted in design;

B.W. performed the measurements and analysis; F.G assisted in simulations; J.C., Y.X, Q.G advised on the optical setup and the interpretation of the data; W.C. and Y.L. supervised the analysis, experiments and edited the manuscript. All authors discussed the results and implications and commented on the manuscript at all stages.

## Additional information

Supplementary Information accompanies this paper.

Competing financial interests: The authors declare no competing financial interests.

# Supplementary Information

## Supplementary Figures

**(Figure S1)**

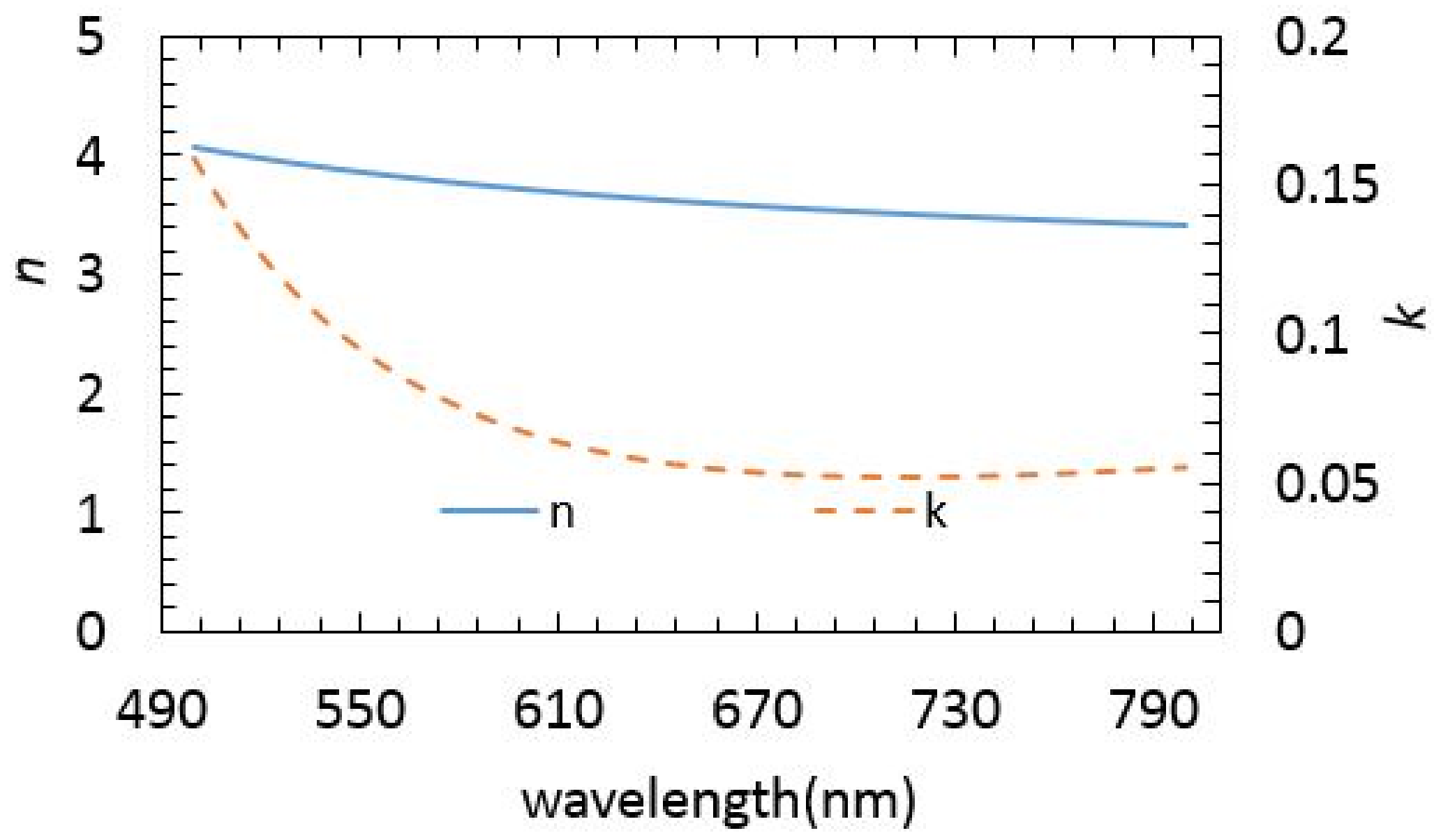

**Supplementary Figure 1 | The optical properties of fabricated 370nm-thick nano-crystalline silicon film.** Experimentally measured complex refractive index of a nano-crystalline silicon film of 370nm thickness in the wavelength range of 500nm-800nm with an ellipsometer.

**(Figure S2)**

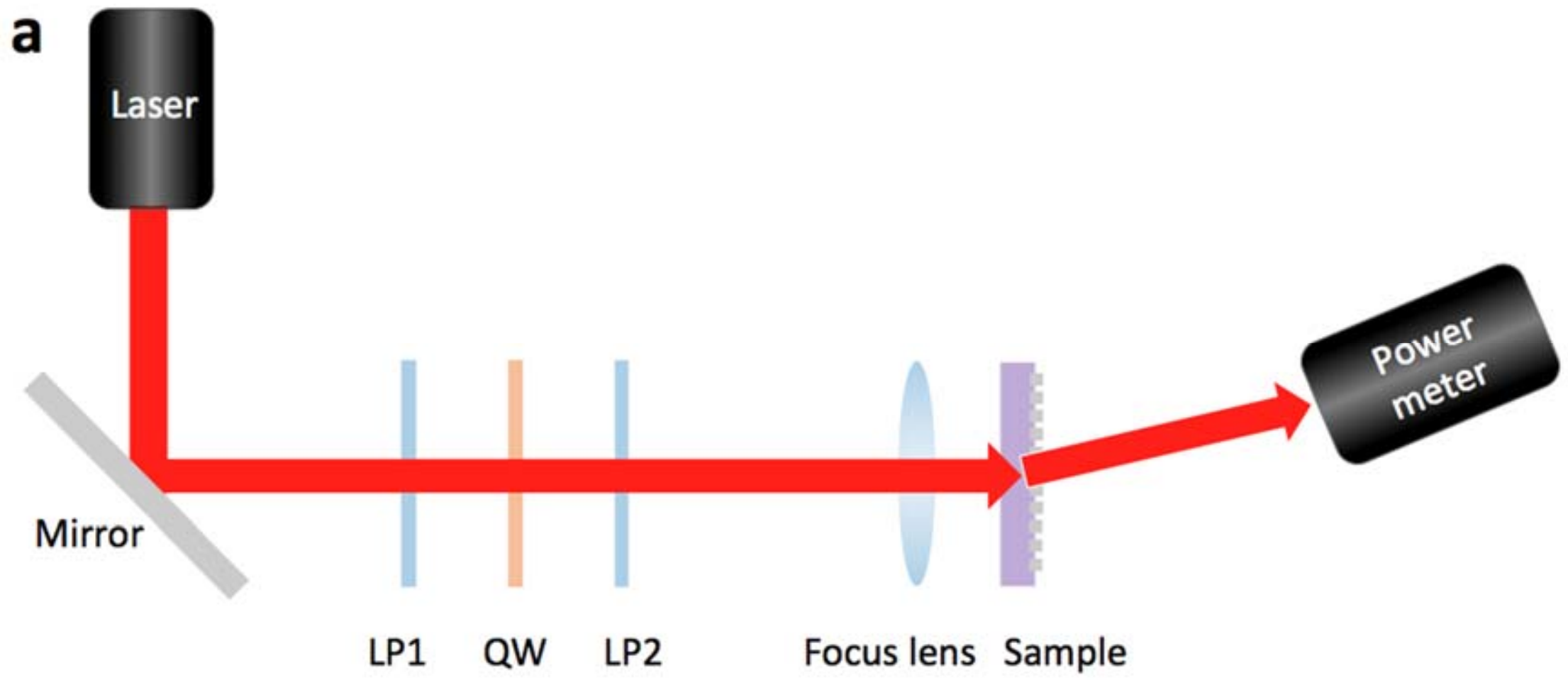

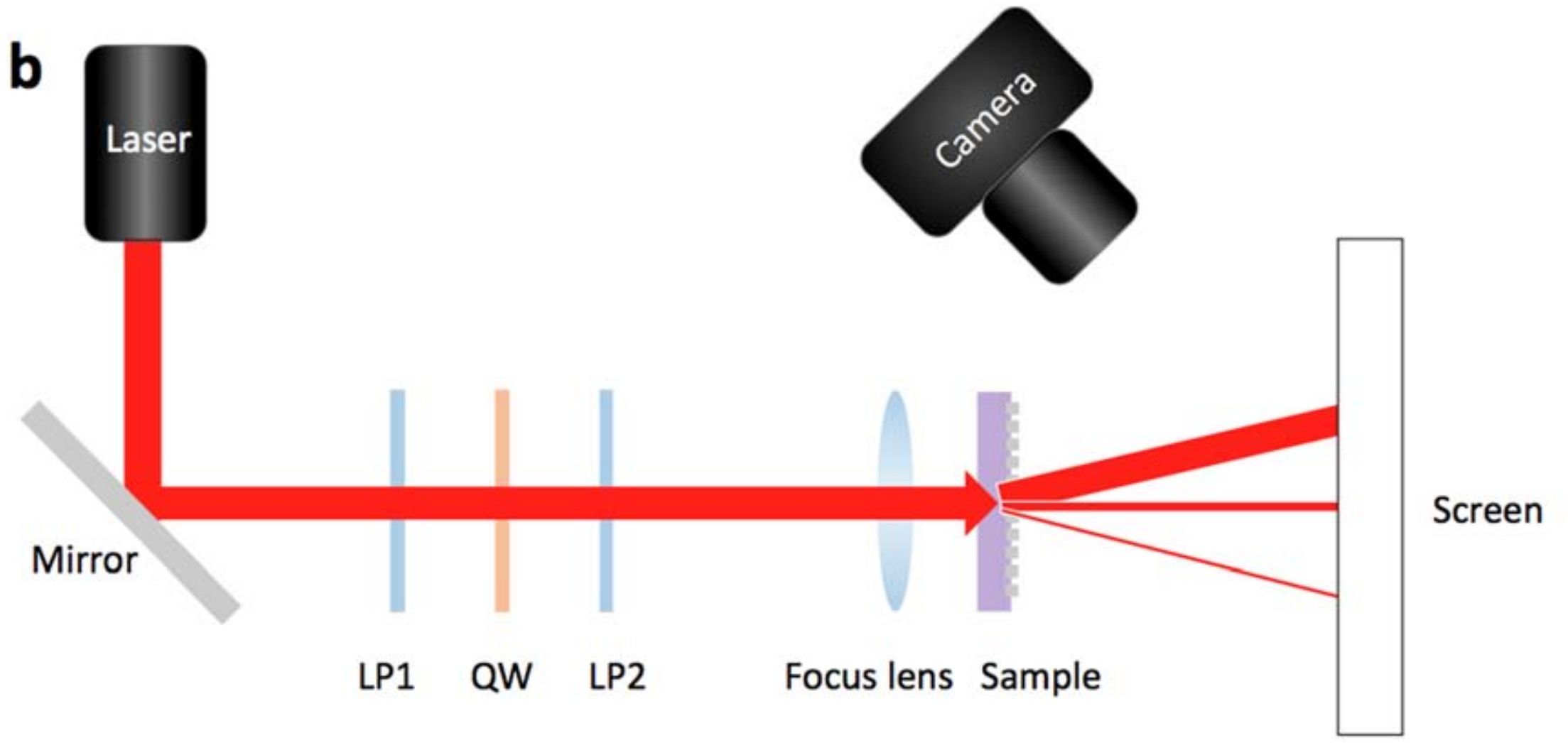

**Supplementary Figure 2 | The schematic of Measurement setup. (a)** The intensity characterization of the transmission light (+1, 0 and -1 order diffraction intensity) using a power meter. **(b)** The light distribution observation behind the gradient metasurface. Abbreviations used for the optical components are: LP for linear polarizer and QW for quarter wave plate.

**(Figure S3)**

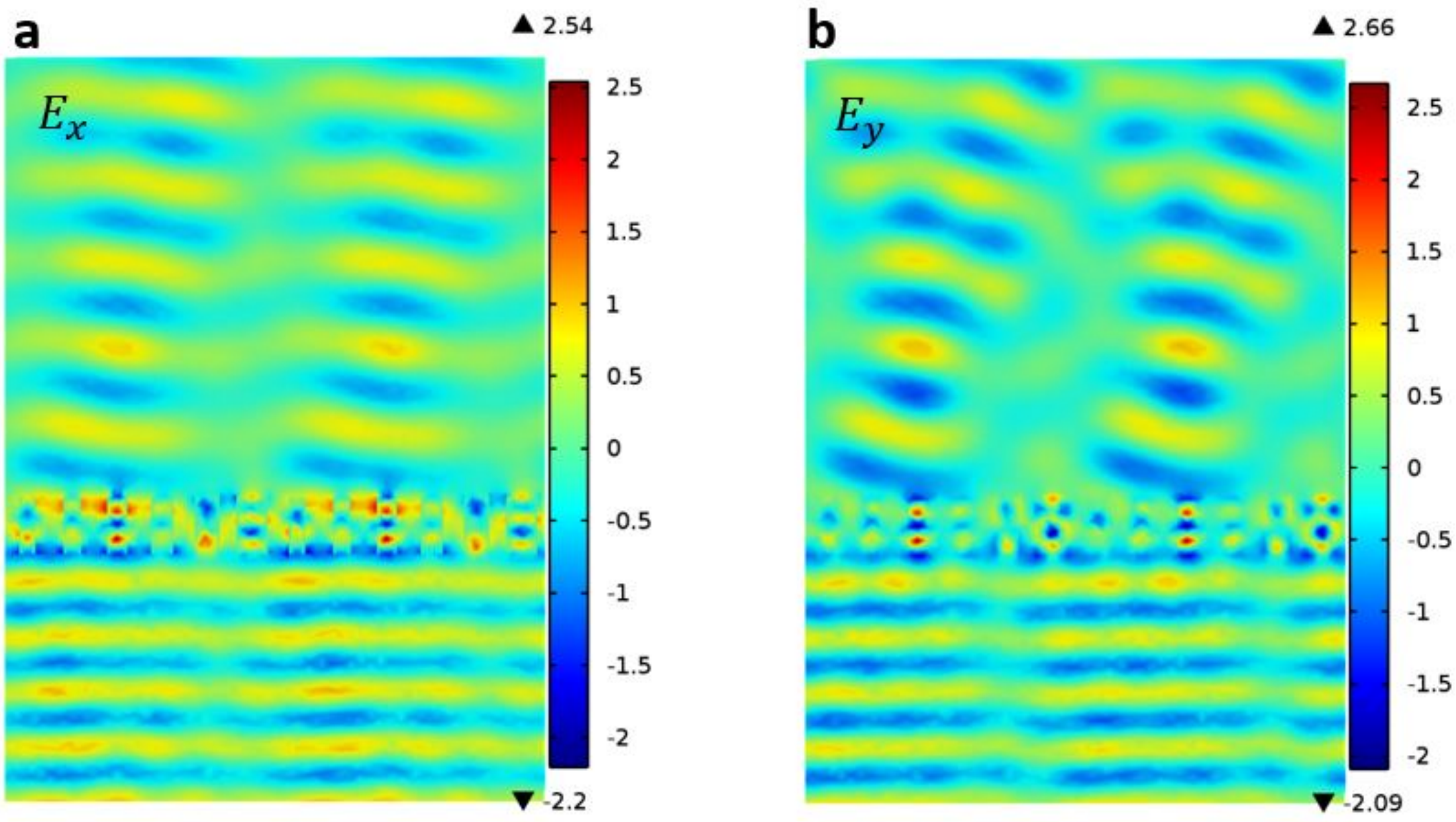

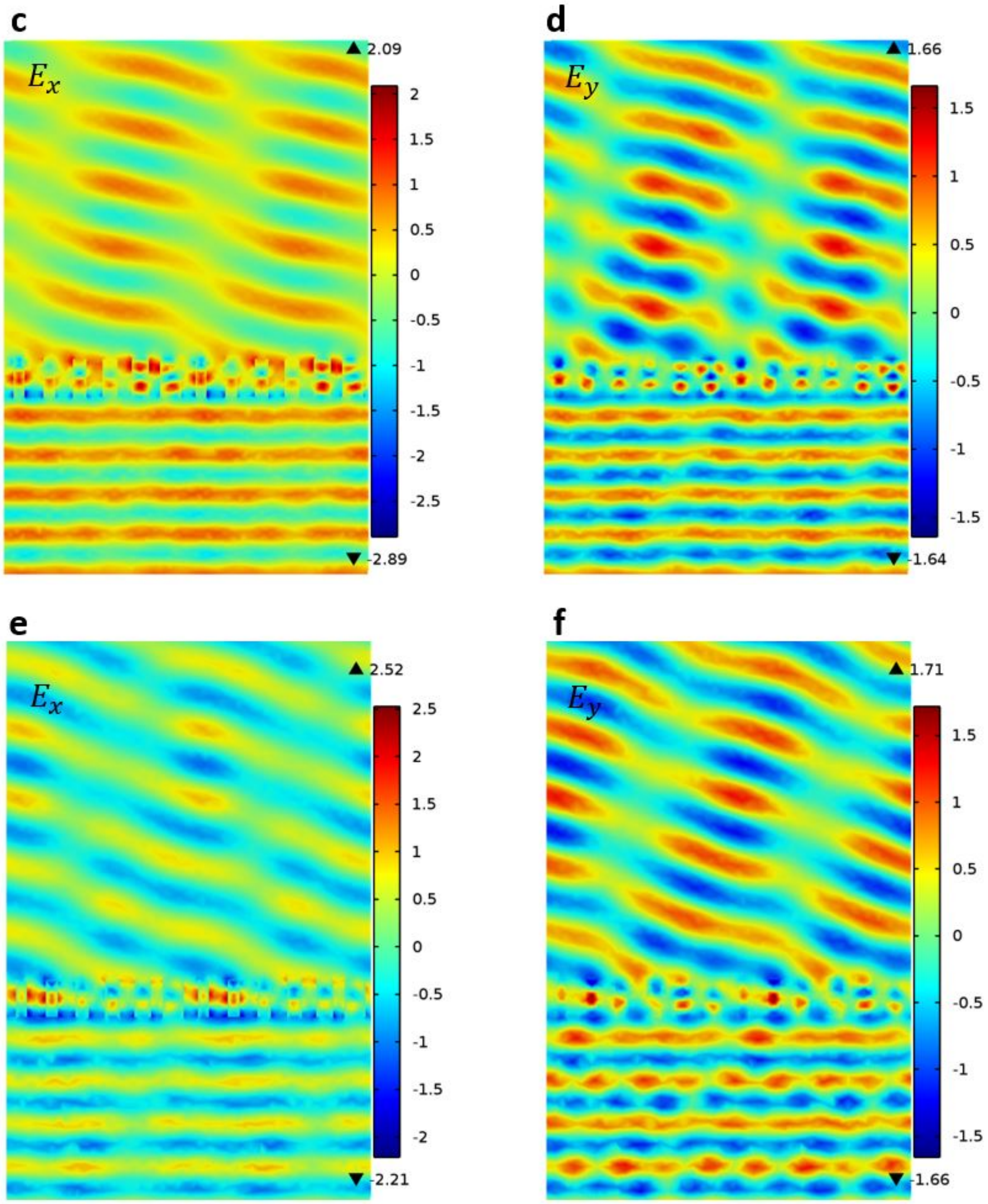

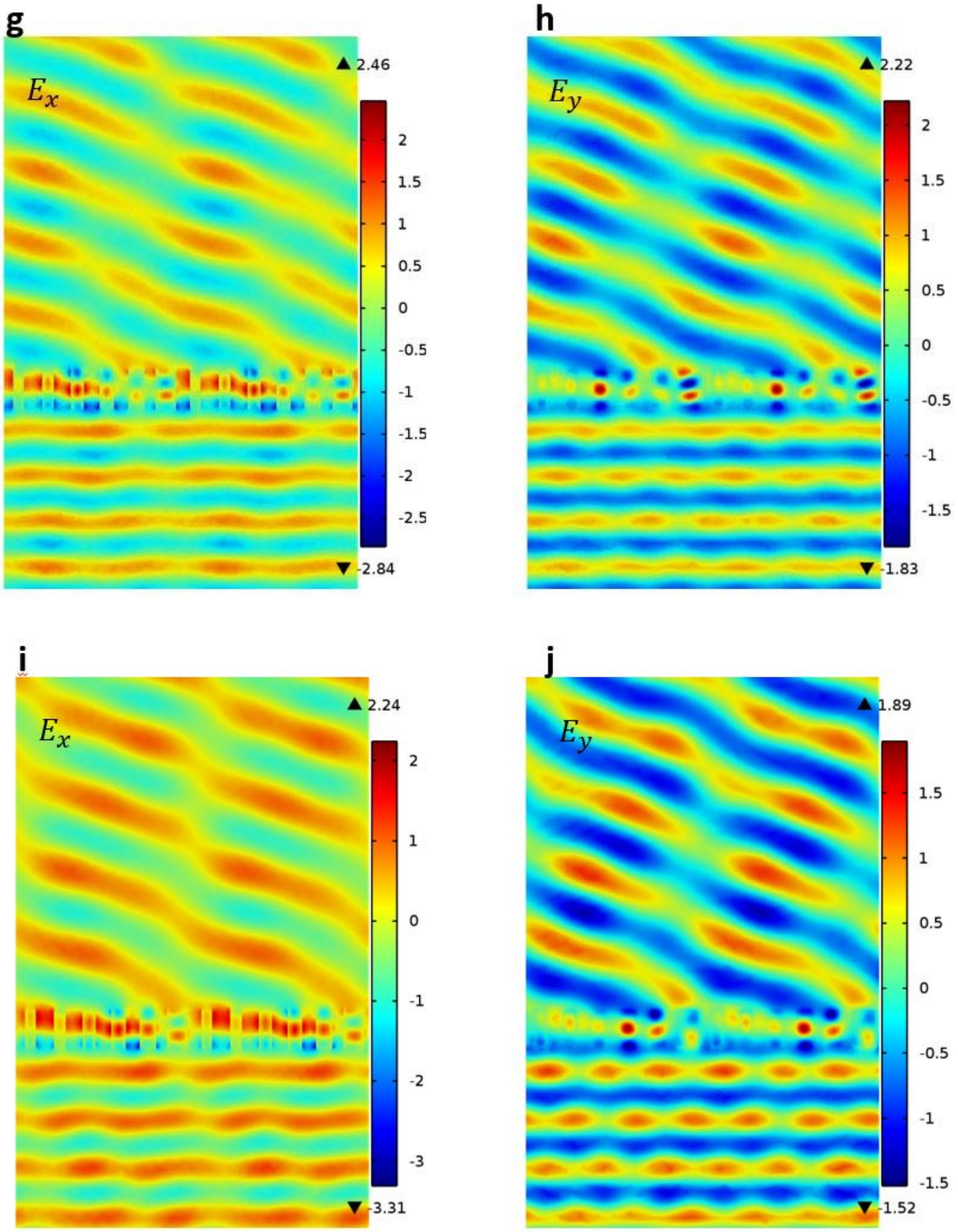

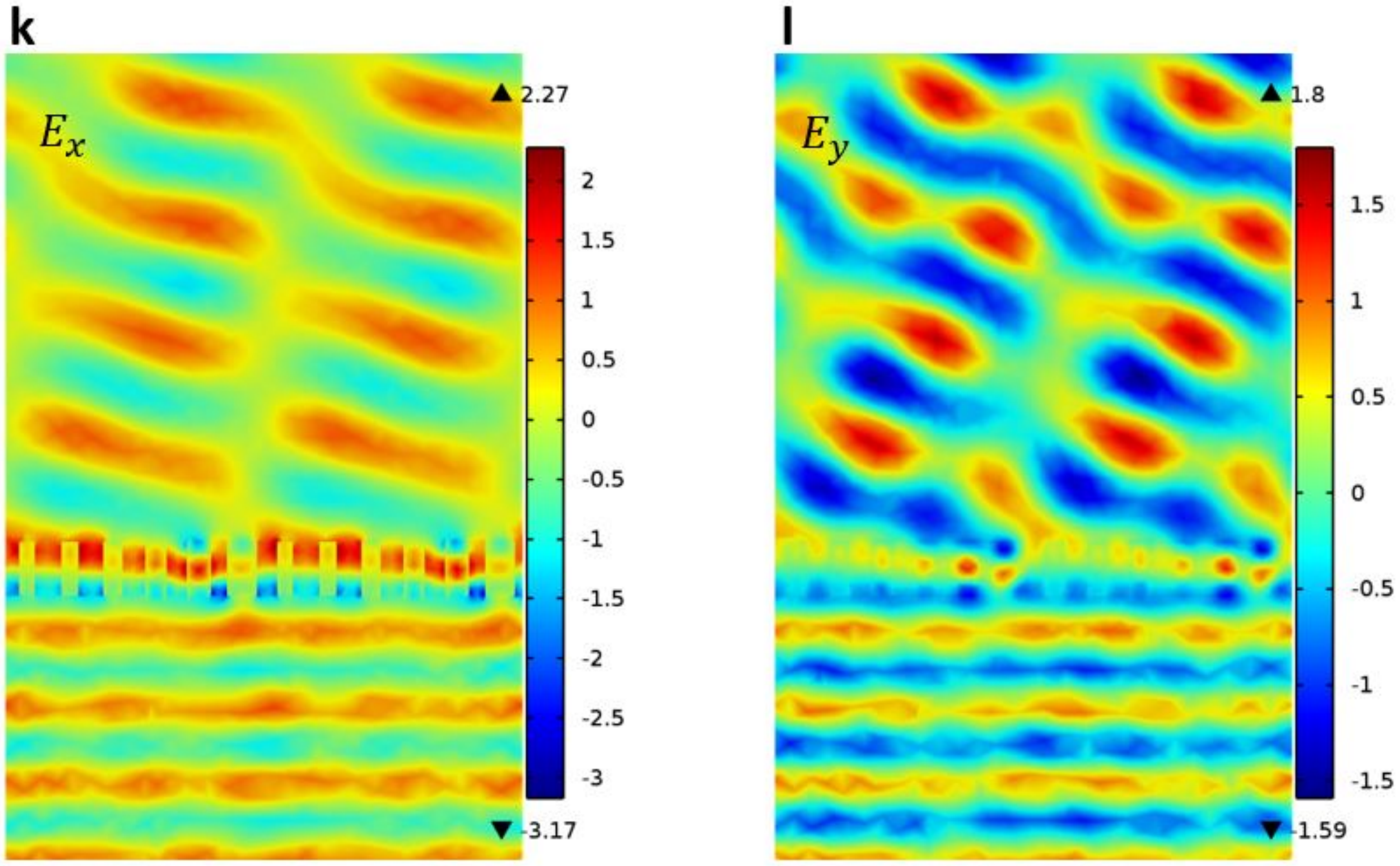

**Supplementary Figure 3 | Full-field simulations.** The transmission electric field patterns of the gradient metasurfaces under the illumination of either $x$-polarized or $y$-polarized incident beam along z-axis at wavelength of (**a**) (**b**) 550nm, (**c**) (**d**) 600nm, (**e**) (**f**) 650nm, (**g**) (**h**) 700nm, (**i**) (**j**) 750nm and (**k**) (**l**) 800nm. The transmission lights deflect evidently from 550nm to 800nm, which have an amazing agreement with the expectations in the main text. Nearly perfect refracted wave fronts are obtained at the wavelength of 650nm and 700nm, which is quite consistent with our experimental results (93% peak efficiency for x polarized incident beam at the wavelength of 670nm and 92% peak efficiency for y-polarized incident beam at the wavelength of 710nm).

# Supplementary Notes

## Supplementary Note 1: Characterization of the deposited nano-crystalline silicon film

The nano-crystalline silicon film is deposited using an ICPECVD system and characterized using a spectroscopic ellipsometer (Sentech SE 850 DUV). The thickness and complex refractive index of the film are obtained by fitting the measured data. The measured complex refractive in the wavelength range from 500nm to 800nm is shown in Fig. S1. The relatively high real part ($n$) makes for a full $2\pi$ phase control and a very low imaginary part ($k$) offers low loss in visible frequencies.

## Supplementary Note 2: Two ways to control the wave fronts

There are two major ways for either plasmonic arrays or dielectric resonators to provide a full $2\pi$ phase delay: the first way is to change the physical geometry of meta-atoms so that the leaky-modes in resonators are tailored and varying phase delay of the scattering light is enabled. The other way, which is based on a purely spatial geometry effect, known as Pancharatnam-Berry phase, is to simply rotate meta-atoms with a circularly polarized incident light, which indicates that the analytical solution for the scattering field never changes in the local coordinates fixed in the meta-atoms. We should note that though the different geometry of meta-atoms significantly influences manipulation efficiency, the second way is genuinely general rather than only feasible to some specific geometry of resonators. Here we choose the first way to control the wave fronts because the concept of Pancharatnam-Berry phase is fundamentally based on polarization status. In fact, it is Pancharatnam-Berry phase that gives rise to the polarization-dependent properties, and our work is to figure out what kinds of symmetries of meta-atoms are immune to that general effect.

**Supplementary Note 3: Verification of the polarization-independent properties**

In the main text, we have demonstrated that for resonators with a multi-fold symmetry ($n$>2), a cross-polarized transmission light cannot be excited in left- and right-circular polarization basis using $\hat{e}_x + i\hat{e}_y$ and $\hat{e}_x - i\hat{e}_y$ as the unit vectors. Utilizing the transmission matrix to express the scattering process by metasurface, we get

$$\mathbf{A}\boldsymbol{E_b}(z) = \boldsymbol{E}(z)$$

Where

$$\mathbf{A} = \mathbf{A}_{\boldsymbol{RL}} = \begin{bmatrix} t_{RR} & 0 \\ 0 & t_{LL} \end{bmatrix}$$

is the transmission matrix with $t_R = \frac{\mathrm{E_R(z)}}{\mathrm{E_0(z)}}$ and $t_L = \frac{\mathrm{E_L(z)}}{\mathrm{E_0(z)}}$.

In the horizontal and vertical polarization frame using $\hat{e}_x$ and $\hat{e}_y$ as the unit vectors, $\mathbf{A}$ should be rewritten as

$$\mathbf{A} = \boldsymbol{A}_{\boldsymbol{xy}} = \begin{bmatrix} t_{xx} & t_{yx} \\ t_{xy} & t_{yy} \end{bmatrix}$$

where $t_{xx} = (\frac{\mathrm{E_R(z)}}{\mathrm{E_0(z)}} + \frac{\mathrm{E_L(z)}}{\mathrm{E_0(z)}})$ , $t_{yx} = \frac{-\mathrm{i}}{2}(\frac{\mathrm{E_R(z)}}{\mathrm{E_0(z)}} - \frac{\mathrm{E_L(z)}}{\mathrm{E_0(z)}})$ , $t_{xy} = (\frac{\mathrm{E_R(z)}}{\mathrm{E_0(z)}} - \frac{\mathrm{E_L(z)}}{\mathrm{E_0(z)}})$ and $t_{yy} = \frac{1}{2}(\frac{\mathrm{E_R(z)}}{\mathrm{E_0(z)}} + \frac{\mathrm{E_L(z)}}{\mathrm{E_0(z)}})$. Taking into account that all transmission matrix can be written as a diagonal matrix using specific $\hat{e}_x$ and $\hat{e}_y$ as the unit vectors[20], $\mathrm{E_R}(z)$ should equal to $\mathrm{E_L}(z)$, and thus transmission matrix can be expressed as

$$\mathbf{A}_{\boldsymbol{xy}} = \begin{bmatrix} t & 0 \\ 0 & t \end{bmatrix} = t\,\mathbf{I}$$

where $t = t_{xx} = t_{yy} = \frac{\mathrm{E_R(z)}}{\mathrm{E_0(z)}}$. In conclusion, the transmission matrix retrogrades to a complex number in the condition that meta-atoms have n-fold symmetry (n>2), which indicates perfect polarization-independent properties.